\documentclass[runningheads]{svmult}
\usepackage{makeidx}   
\usepackage{graphicx}  
\usepackage{subeqnar}  
\usepackage{multicol}  
\usepackage{physprbb}  
\makeindex             
\begin{document}
\title*{Open charm production in deep inelastic 
\protect\newline scattering at next-to-leading order at HERA
\footnote{Invited lecture presented at 
{\em Ringberg Workshop: New Results from HERA}, 
30 May - 4 June 1999, Schlo\ss{} Ringberg,
Rottach-Egern, Germany, to appear in proceedings thereof}}
\toctitle{Open charm production in deep inelastic
\protect\newline scattering at next-to-leading order at HERA}
%
%
\titlerunning{Open charm production in deep inelastic
scattering \ldots}
%
\author{Brian W. Harris}
\authorrunning{B. W. Harris}
%
%
\institute{HEP Theory Group, Argonne National Laboratory, 
           Argonne, IL 60439, USA}

\maketitle              

\begin{abstract}
An introduction and overview of charm production in deep inelastic 
scattering at HERA is given.  The existing next-to-leading order perturbative 
QCD calculations are then reviewed, and key results are summarized.  Finally, 
comparisons are made with the most recent HERA data, and unresolved 
issues are highlighted.
\end{abstract}

\vspace*{-7.3cm} \noindent ANL-HEP-CP-99-69
\vspace*{ 6.4cm}

\section{Introduction}

Electromagnetic interactions have long been used to study both 
hadronic structure and strong interaction dynamics.  Examples include 
deep inelastic lepton-nucleon scattering, hadroproduction of 
lepton pairs, the production of photons with large 
transverse momenta, and various photoproduction processes involving 
scattering of real or very low mass virtual photons from hadrons.  
In particular, heavy quark production in deep inelastic 
electron-proton scattering (DIS) is calculable in QCD and 
provides information on the gluonic content of the proton which 
is complementary to that obtained in direct photon production or 
structure function scaling violation measurements.  In addition, the scale 
of the hard scattering may be large relative to the 
mass of the charm quark, thus allowing one to study whether and 
when to treat the charm as a parton.

Early measurements of open charm production in neutral current DIS, 
performed by the Berkeley-Fermilab-Princeton (BFP) \cite{bfp} 
and European Muon Collaboration (EMC) \cite{emc} experiments, 
touched upon the topics relevant for HERA today:  
production mechanism, charm fragmentation, gluon parton distribution function 
extraction, and charm contribution to the proton structure function.
See \cite{strovink} for a review of these experiments.
Activity in this area has increased recently with new data available 
from the H1 \cite{h1} and ZEUS \cite{zeus} experiments at HERA.  In 
particular, substantial samples of reconstructed $D^*$ hadrons have 
been obtained, and a semi-lepton decay mode analysis is 
underway\cite{wouter}.  Considerably more data is anticipated in the 
next few years.

Interest in the production mechanism is twofold.
First, one is concerned with the leading twist-two term in 
the operator product expansion which incorporates the factorization 
theorem for hard scattering.
Second, there is interest in studies of a higher twist 
charm component to the proton.

When considering the leading twist-two term the main issue is whether 
and when to treat the charm as a parton.  Near threshold it is well 
established that charm is produced through photon-gluon fusion.  
On the other hand, very far above threshold the charm should be 
resummed into an effective parton distribution.  How one interpolates 
from one region to another is described by a variable flavor number 
scheme, several of which have been proposed recently \cite{var,buzab,tr}.  
One must look at sufficiently inclusive observables in order to 
build up the logarithms that are to be 
summed, so predictive power for some observables is lost.
In other words, for differential quantities, the fixed flavor number 
scheme of photon-gluon fusion provides the most appropriate 
formalism \cite{cs}.

The idea of a higher twist charm component to the proton was 
introduced by Brodsky {\em et al}.\ \cite{ic}.
In this scenario the intrinsic charm quark Fock component in the 
proton wave function, $|uudc\bar{c}>$, is generated by virtual 
interactions where gluons couple to two or more valence quarks.  
The probability for $c\bar{c}$ fluctuations to exist in a light hadron 
scales as ${\cal O}(1/m_c^2)$ relative to the twist-two 
component, and is therefore higher twist.  
The EMC data \cite{emc} marginally support this idea.  
A full re-analysis of the EMC data has 
been carried out including both leading twist and intrinsic components 
at NLO including mass and threshold effects \cite{hsv}.  
The result is that one data point contributes to a $0.8\pm0.6\%$ 
normalization of the intrinsic component relative to leading twist.

Fragmentation is the most contentious topic at present.  The 
production and subsequent hadronization of charmed quarks in DIS is 
not as clean as the much studied case of production in $e^+e^-$ 
annihilation.  In particular, in DIS there are proton beam remnants 
around which necessarily talk to the charm quark as it hadronizes.  
If events too close to the beam direction are selected, one can 
expect deviations from models which do not account for this effect.
Presently it is hoped that detailed experimental 
studies in the Breit frame, wherein one hemisphere resembles $e^+e^-$, 
will provide additional information on the fragmentation process.

Because the production method is dominated by the leading twist 
photon-gluon fusion near threshold, it is possible to extract a 
gluon parton distribution function 
(PDF) and compare with existing PDFs, which derive their gluon information 
from other sources, such as direct photon production or 
structure function scaling violation measurements.
This has been done by both H1 and ZEUS collaborations.  
The results are completely 
consistent with comparable PDF sets in the same $x$ and $Q^2$ range.  
While unlikely to ever replace structure function scaling violation 
measurements, charm production does serve as a nice consistency check on the 
gluon PDF.

One may also wish to take the open charm measurement and extrapolate the 
cross section over the full phase space and then extract the charm's 
contribution to the proton structure function.  
Historically the structure functions 
have been useful as input into global analyses and a testing ground for the 
variable flavor number schemes mentioned above.  
Care must be taken so that conclusions drawn are not artifacts of 
the extrapolation procedure (model).

\begin{figure}
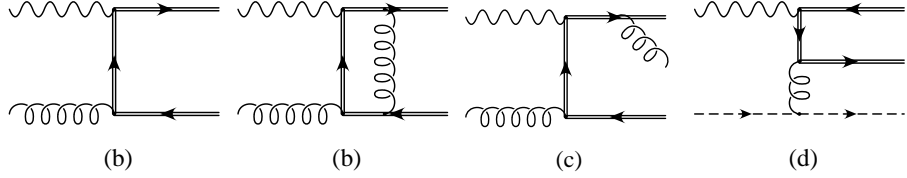

\begin{center}
\includegraphics[width=.24\textwidth]{harris.1.1}
\includegraphics[width=.24\textwidth]{harris.1.2}
\includegraphics[width=.24\textwidth]{harris.1.3}
\includegraphics[width=.24\textwidth]{harris.1.4}
\end{center}
\caption[]{Typical Feynman diagrams contributing to the amplitude for 
neutral current charm production.  (\textbf{a}) An order $eg_s$ Born diagram.  
(\textbf{b}) An order $eg_s^3$ virtual diagram.  (\textbf{c}) An order 
$eg_s^2$ gluon-bremsstrahlung diagram.  (\textbf{d}) An order $eg_s^2$ light 
quark initiated diagram}
\label{fig1}
\end{figure}

\section{Next-to-leading Order Calculations}
\label{sec2}

The reaction under consideration is charm quark production via 
neutral-current electron-proton scattering, 
$e^{-}(l) + P(p) \rightarrow e^{-}(l') + c(p_1) + X\,$.
When the momentum transfer squared 
$Q^2=-q^2>0$ ($q=l-l'$) is not too large $Q^2 \ll M_Z^2$, 
the contribution from $Z$ boson exchange is suppressed 
and the process is dominated by virtual-photon exchange.  
After an azimuthal integration, the cross section may be written 
in terms of structure functions $F^c_2(x,Q^2,m_c)$ and $F^c_L(x,Q^2,m_c)$ 
as follows:
\begin{equation}
\label{cross}
\frac{d^2\sigma}{dydQ^2} 
= \frac{2\pi\alpha^2}{yQ^4} \left\{ \left[ 1 + (1-y)^2 \right] 
F^c_2(x,Q^2,m_c) - y^2 F^c_L(x,Q^2,m_c) \right\}
\end{equation}
where $x = Q^2 / 2p \cdot q$ and $y = p\cdot q / p \cdot l$ 
are the usual Bjorken scaling variables, $\alpha$ is
the electromagnetic coupling, and $m_c$ is the charm mass.  
The scaling variables are related to the 
square of the center of momentum energy of the electron-proton 
system $S=(l+p)^2$ via $xyS=Q^2$.  
The total cross section is given by \cite{schuler}
\begin{equation}
\label{cross2}
\sigma = \int_{4m_c^2/S}^1 dy \int_{m_e^2 y^2/(1-y)}^{yS-4m_c^2} dQ^2
\left( \frac{d^2\sigma}{dydQ^2} \right)
\end{equation}
where $m_e$ is the electron mass.  

Typical Feynman diagrams for this process are shown in Fig.~\ref{fig1}.  
The interference of the ${\cal O}(eg_s)$ Born diagrams with 
the ${\cal O}(eg_s^3)$ one-loop virtual diagrams 
produces a result that is ${\cal O}(\alpha\alpha_s^2)$.  
The ultraviolet divergences are removed by 
renormalization in the Collins-Wilczek-Zee scheme \cite{cwz}.  
The result is added to the square of the gluon-bremsstrahlung diagrams 
which is also ${\cal O}(\alpha\alpha_s^2)$.  Initial state collinear 
singularities are mass factorized to obtain a finite NLO result.
For the light quark initiated subprocess there are no virtual contributions 
at ${\cal O}(\alpha\alpha_s^2)$.  One only encounters 
initial state collinear singularities which are again removed 
using factorization.  The full NLO corrections were first calculated 
in \cite{lrsvn1} and may be written in the form 
\vspace{4ex}
\begin{eqnarray}
\label{fhad}
F_k^c(x,Q^2,m_c) &=& 
\frac{Q^2 \alpha_s(\mu^2)}{4\pi^2 m_c^2}
\int_{\xi_{\rm min}}^1 \frac{d\xi}{\xi} \Bigl\{ 
e_c^2 f_{g/P}(\xi,\mu^2) 
\nonumber \\ &&  \times
\Bigl[ c^{(0)}_{k,g} + 4 \pi \alpha_s(\mu^2) 
\Bigl( c^{(1)}_{k,g} + \bar c^{(1)}_{k,g} \ln \frac{\mu^2}{m_c^2} \Bigr)
\Bigr]  \nonumber \\
&+&  4 \pi \alpha_s(\mu^2) \sum_{i=q,\bar q} f_{i/P}(\xi,\mu^2)
\nonumber \\ && \times
\Bigl[ e_c^2 \Bigl( c^{(1)}_{k,i} + \bar c^{(1)}_{k,i} \ln \frac{\mu^2}{m_c^2}
\Bigr) 
+  e^2_i \, d^{(1)}_{k,i} + e_i\, e_c \, o^{(1)}_{k,i} \,
\Bigr] \Bigr\} \nonumber \\ &&
\end{eqnarray}
with $k = 2,L$.
The lower limit on the integration over the parton momentum fraction $\xi$ 
is $\xi_{\rm min} = x(4m_c^2+Q^2)/Q^2$.
The parton momentum distributions in the proton are denoted by
$f_{i/P}(\xi,\mu^2)$.  The sum is taken over the light quarks, $q=u,d,s$.
The mass factorization scale $\mu_f$ has been 
set equal to the renormalization scale $\mu_r$ and is denoted by $\mu$.
All charges are in units of $e$. 
$c^{(0)}_{k,i}$, $c^{(1)}_{k,i}$, 
$\overline c^{(1)}_{k,i}\,, (i = g, q, \bar q)$,  and 
$d^{(1)}_{k,i} $, $o^{(1)}_{k,i} \,,(i= q, \bar q)$ are scale 
independent parton coefficient functions, and are distinguished by 
their origin.  The $c$-coefficient functions
originate from processes involving the virtual photon-heavy quark coupling, 
the $d$-coefficient functions arise from processes involving the
virtual photon-light quark coupling, and the $o$-coefficient functions are
from the interference between these processes.

In addition to describing the structure functions,
the coefficient functions of \cite{lrsvn1} may also be used to 
compute the single-inclusive distributions $dF_k/dp_t$ and $dF_k/dy$ 
\cite{lrsvn2}, 
where $p_t$ and $y$ are the transverse momentum and rapidity, respectively, 
of the heavy quark in the virtual photon-proton center of momentum system.

Unfortunately, the analytic expressions of the coefficient functions 
are too long to be published in journal form.  They are however available as 
computer code, but, for the same reason, the code tends to be slow.  
Initially this was seen as an impediment to including them in a global 
fitting program.  Then in \cite{rsvn}, with the threshold and asymptotic 
behavior of the coefficient functions removed, it was possible to 
numerically tabulate grids, with a fast interpolation routine, 
so that speedy computation of Eq.\ (\ref{cross}) -- (\ref{fhad}) 
became possible.  This process was further refined in \cite{work1}.

The coefficient functions in the large momentum transfer limit are a 
necessary ingredient for constructing the variable flavor number schemes 
mentioned in the introduction.  Exact analytic formulae for the 
$d^{(1)}_{k,i}$ together with analytic formulae for the 
coefficient functions $c^{(0)}_{k,i}$, $c^{(1)}_{k,i}$, 
$\overline c^{(1)}_{k,i}\,, (i = g, q, \bar q)$,  
and $d^{(1)}_{k,i}\,, (i= q, \bar q)$ in the limit $Q^2 \gg m^2$ 
can be found in \cite{buzaa}.

The coefficient functions have also been calculated in a fully 
differential form \cite{hs1}.  These in turn can be used with 
Eq.\ (\ref{fhad}) to construct pair-inclusive distributions such 
as $dF_k/dM_{c\bar{c}}$ \cite{hs2} where $M_{c\bar{c}}$ is the 
invariant mass of the produced charm-anticharm system.

The resulting differential structure functions and Eq.\ (\ref{cross}) 
have further been used to construct a NLO monte 
carlo style program, {\sc HVQDIS} \cite{bh,hs3}.
The basic components (in terms of virtual-photon-proton scattering) 
are the 2 to 2 body squared matrix elements through one-loop order and 
tree level 2 to 3 body squared matrix elements, 
for both photon-gluon and photon-light-quark 
initiated subprocesses, as shown in Fig.~\ref{fig1}.  
It is therefore possible to study fully-, single-, and semi-inclusive 
production at NLO, and three body final states 
at leading order.
The goal of this NLO calculation 
is to organize the soft and collinear singularity 
cancellations without loss of information 
in terms of observables that can be predicted.

The subtraction method provides a mechanism for this cancellation.
It allows one to isolate the soft and collinear singularities 
of the 2 to 3 body processes within the framework of dimensional 
regularization without calculating all the phase space integrals in 
a space-time dimension $n\ne 4$.
Expressions for the three-body squared matrix elements in the limit 
where an emitted gluon is soft appear in a factorized form where 
poles $\epsilon=2-n/2$ multiply leading order squared matrix elements.  
These soft singularities cancel upon addition 
of the interference of the leading order diagrams with the 
renormalized one-loop virtual diagrams.
The remaining singularities are initial state collinear in origin. 
The three-body squared matrix elements appear in a 
factorized form, where poles in $\epsilon$ multiply splitting 
functions convolved with leading order squared matrix elements. 
These collinear singularities are removed through mass 
factorization.

The result of this calculation is an expression 
that is finite in four-dimensional space time.  One can 
compute all phase space integrations using 
standard monte carlo integration techniques.
The final result is a program which returns 
parton kinematic configurations and their corresponding 
weights, accurate to ${\cal O}(\alpha\alpha_s^2)$.
The user is free to histogram any set of 
infrared-safe observables and apply 
cuts, all in a single histogramming subroutine.
Additionally, one may study heavy hadrons using the  
Peterson {\em et al}.\ model \cite{pete}.
Detailed physics results from this program 
and a description of the necessary cross checks 
the program satisfies are given in \cite{hs3}.
See also \cite{work3} for recent improvements.

\begin{figure}
\begin{center}
\includegraphics[width=0.9\textwidth]{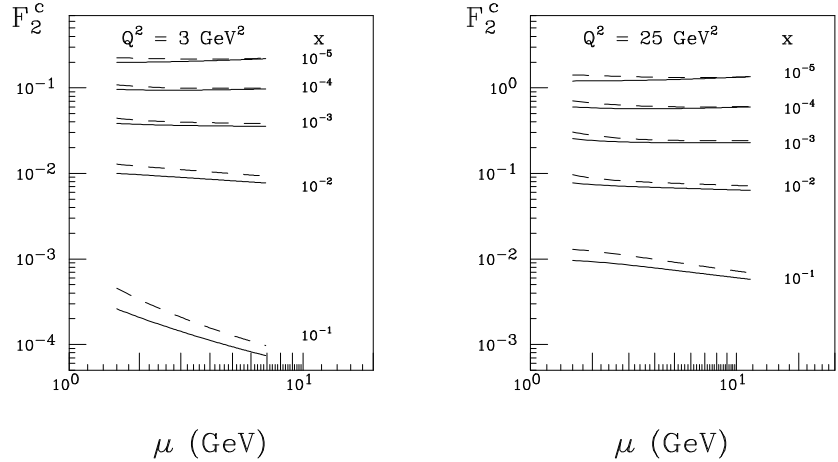}
\end{center}
\caption[]{The renormalization/factorization scale dependence of the 
structure function $F_2^c(x,Q^2,m_c)$ 
for $Q^2 = 3$ GeV$^2$ (left) and $Q^2 = 25$ GeV$^2$ (right) for various 
$x$ values.  The results for GRV94 ({\it solid lines\,}) and CTEQ4F3 
({\it dashed lines\,}) parton distribution sets are shown}
\label{fig2}
\end{figure}

\section{Results and Current Issues}
\label{sec3}

Charmed meson differential cross sections are measured 
experimentally \cite{bfp,emc,h1,zeus} within some detector 
acceptance region, and the corresponding theory predictions can 
be made using {\sc HVQDIS}.
As we saw above, the cross section is an integral over the 
structure functions.  
Therefore, the two share many of the same properties.  
Before discussing the cross sections, the more salient features of the 
NLO structure functions will be reviewed.
The interested reader can find additional details in the original paper 
\cite{lrsvn1} and, more so, in the recent 
phenomenological analyses \cite{work1,grs,v96,work2}.  

For moderate $Q^2 \sim 10\,{\rm GeV}^2$ one finds that the charm quark 
contribution at small $x \sim 10^{-4}$ is approximately $25\%$ of the 
total structure function (defined as light parton plus heavy quark 
contributions).  In contrast, the contribution from bottom 
quarks is only a few percent due to charge and phase space suppression. 
The structure functions show a marked rise at small $x$ 
due primarily to the rapidly rising gluon distribution:
the gluon initiated contributions comprise most of the structure function.  
The light quark initiated processes give only a few percent contribution 
at small $x$.  
The scale dependence of the structure functions is very small in the 
HERA $x$ and $Q^2$ regions.  
This is demonstrated in Fig.~\ref{fig2} for various $x$ and $Q^2$ values.
The largest variation comes from our imprecise knowledge 
of the charm quark mass.  For example, a $\pm 10\%$ variation of the 
charm mass about the central value of $1.5$ GeV gives a $\pm 20\%$ 
variation in the structure function for small $x$ and moderate $Q^2$.

At moderate $Q^2$, and $x$ values larger than $0.01$, the charm 
structure function is increasingly dominated by
partonic processes near the charm quark pair-production threshold. 
The large size of the gluon density for small momentum 
fractions gives relatively large weight to such processes \cite{v96}. 
Although the QCD corrections at presently accessible $x$ values are 
moderate (about $30-40\%$),
with an increasing amount of data to be gathered at higher $x$,
it is worthwhile to have a closer look at such near-threshold subprocesses.
In this kinematic region, the QCD 
corrections are dominated by large Sudakov double logarithms.
Recently \cite{lm99}, these Sudakov logarithms have been resummed to 
all orders of perturbation theory, to next-to-leading logarithmic 
accuracy, and, moreover, in single-particle inclusive kinematics 
\cite{los98}. Let us recall the main results. 
First, the quality of the approximation for the {\it next-to-}leading 
logarithmic threshold resummation was found to be clearly superior to 
leading logarithmic one.  Furthermore, the resummation provided 
next-to-next-to-leading order estimates \cite{lm99}, which were found 
to be sizable for $x\geq 0.05$.

\begin{figure}[t]
\begin{center}
\includegraphics[width=0.8\textwidth]{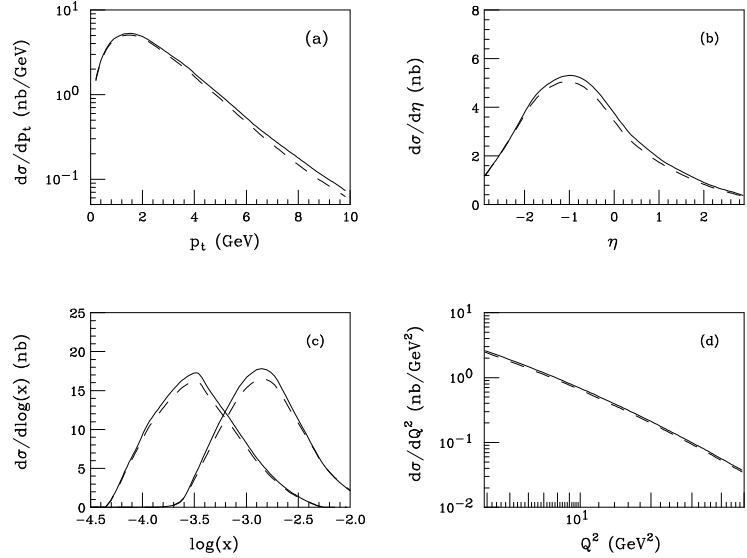}
\end{center}
\caption[]{Next-to-leading order differential cross sections for charm 
production at $\sqrt{S}=301$ GeV at HERA using the GRV94 ({\it dashed 
lines\,}) and CTEQ4F3 ({\it solid lines\,}) parton distribution sets.  
(\textbf{a}) Transverse momentum.
(\textbf{b}) Pseudo-rapidity.
(\textbf{c}) Bjorken $x$ ({\it left set\,}) and parton momentum 
fraction $\xi$ ({\it right set\,}).  (\textbf{d}) 
Photon virtuality $Q^2$}
\label{fig3}
\end{figure}

Let us now examine some of the properties of the charm quark cross section.
Results are presented in the HERA lab frame with positive rapidity in the 
proton direction.  The proton and electron beam energies are taken to be 
$820\, {\rm GeV}$ and $27.6\, {\rm GeV}$, respectively.  Results are 
for the kinematic range $3<Q^2<50\, {\rm GeV}^2$ and $0.1<y<0.7$.
The CTEQ4F3 \cite{cteq4f3} and GRV94 HO \cite{grv94} proton-parton 
distribution sets are used. The renormalization and factorization scales 
have been set equal to $\mu$.

Fig.~\ref{fig3} shows the NLO cross 
sections differential in transverse momentum $p_t$, pseudo-rapidity $\eta$, 
Bjorken $x$, and momentum transfer $Q^2$ for charm quark production using 
the GRV94 (dashed) and CTEQ4F3 (solid) parton 
distribution sets at $\mu=\protect\sqrt{Q^2+4m_c^2}$ with $m_c=1.5\, 
{\rm GeV}$.  From Eq.\ (\ref{fhad}) the parton distributions are 
probed at a momentum fraction $\xi$ which is typically one order of 
magnitude larger the 
$x$.  This is illustrated in Fig.~\ref{fig3}c where a plot 
of $d\sigma/d\log(\xi)$ vs.\ $\log(\xi)$ (right set of curves) 
is superimposed on the plot of 
$d\sigma/d\log(x)$ vs.\ $\log(x)$ (left set of curves).  
The difference between the curves is approximately 
$10\%$ at $\xi = 10^{-2.7}$.

\begin{figure}[t]
\begin{center}
\includegraphics[width=0.8\textwidth]{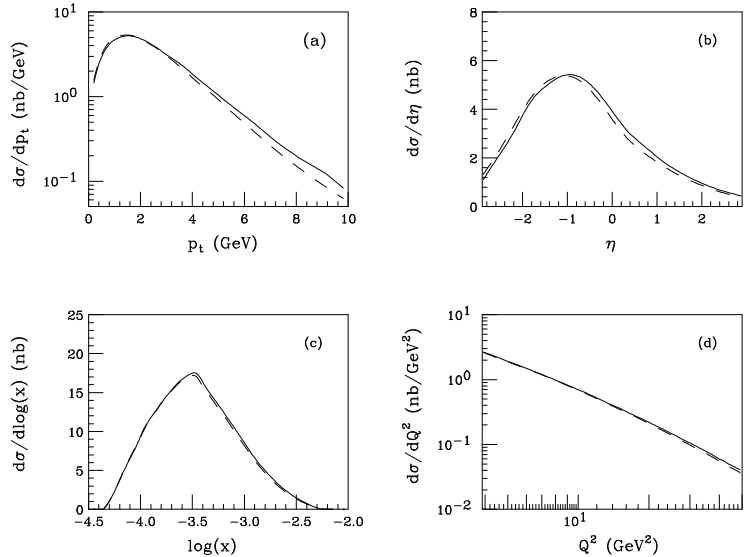}
\end{center}
\caption[]{Same set of distributions as Fig.~\ref{fig3}, but this time 
showing the variation with respect to renormalization/factorization 
scale, $\mu=2m_c$ ({\it solid lines\,}) and $\mu=2\sqrt{Q^2+4m_c^2}$ 
({\it dashed lines\,})}
\label{fig4}

\end{figure}
\begin{figure}
\begin{center}
\includegraphics[width=0.8\textwidth]{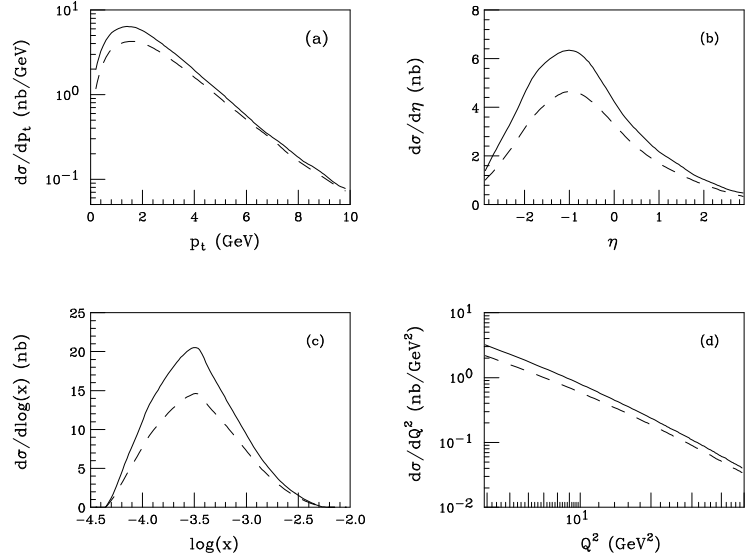}
\end{center}
\caption[]{Same set of distributions as Fig.~\ref{fig3}, but this time 
showing the variation with respect to charm quark mass, 
$m_c=1.35$ GeV ({\it solid lines\,}) and $m_c=1.65$ GeV 
({\it dashed lines\,})}
\label{fig5}
\end{figure}

The scale dependence of the NLO differential cross sections is  
shown in Fig.~\ref{fig4}.  The curves were made using the 
CTEQ4F3 parton distribution 
set at $\mu=2m_c$ (solid) and $\mu=2\sqrt{Q^2+4m_c^2}$ (dashed) with 
$m_c=1.5\, {\rm GeV}$.  The curves show very little scale dependence.  
This can be anticipated from the results shown in Fig.~\ref{fig2} 
and the distribution in Bjorken $x$ shown in Fig.~\ref{fig4}c.  
The latter shows the cross section is dominated by 
$x \sim 10^{-3.5}=3.2 \times 10^{-4}$ while the former shows that, 
independent of $Q^2$, the structure function is very flat in this particular 
$x$ region.  Therefore, the cross section tends to be fairly insensitive 
to the choice of scale.
Other kinematic regions show increased scale dependence.  

\begin{figure}
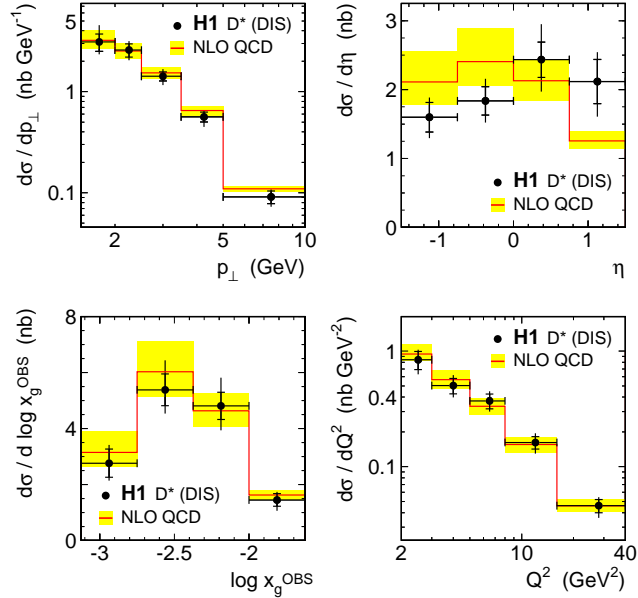

\begin{center}
\includegraphics[width=.34\textwidth]{harris.6.1}
\includegraphics[width=.34\textwidth]{harris.6.2}
\end{center}
\begin{center}
\includegraphics[width=.34\textwidth]{harris.6.3}
\includegraphics[width=.34\textwidth]{harris.6.4}
\end{center}
\caption[]{Various differential cross sections for $D^*$ meson production 
at HERA as measured by the H1 collaboration \cite{h1} compared 
to the next-to-leading order calculation described in the Sec.~\ref{sec2} 
plus a Peterson {\it et al.} fragmentation model.
The shaded band corresponds to varying 
the charm quark mass from $1.3$ to $1.7$ GeV}
\label{fig6}
\end{figure}

The largest uncertainty in the structure function calculation is due  
to the charm quark mass.  The same is true for the cross section 
as shown in Fig.~\ref{fig5}.  The NLO differential cross 
sections for charm quark production using the CTEQ4F3 parton distribution 
set at $\mu=\sqrt{Q^2+4m_c^2}$ with $m_c=1.35\, {\rm GeV}$ (solid) and 
$m_c=1.65\, {\rm GeV}$ (dash) are shown.
Mass effects are smaller at the larger transverse mass because they are 
suppressed by powers of $m_c/p_t$ in the matrix elements.  
However, if the range is extended much further, large logarithms of the form 
$\ln(p_t^2/m_c^2)$ appear in the cross section and should be resummed.

\begin{figure}
\begin{center}
\includegraphics[width=0.8\textwidth]{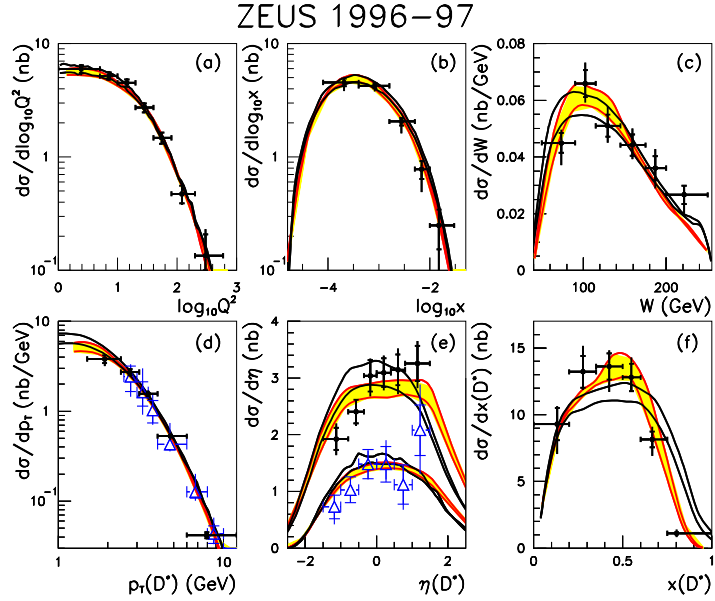}
\end{center}
\caption[]{Various differential cross sections for $D^*$ meson production 
at HERA as measured by the ZEUS collaboration \cite{zeus} compared 
to the next-to-leading order calculation described in the Sec.~\ref{sec2} 
plus a Peterson {\it et al.} fragmentation model.   For the shaded 
band the Peterson {\it et al.} model was replaced by an effective 
fragmentation model extracted from RAPGAP which includes  
a drag effect between the proton remnant and the hadronizing charm.
The bands result from varying the charm quark mass from $1.3$ to $1.5$ GeV}
\label{fig7}
\end{figure}

Before closing, we compare the NLO calculations described 
above with the most recent HERA data from H1 \cite{h1} and ZEUS \cite{zeus} 
collaborations.  
The measurements make use of a tagging technique wherein the $D^*$ meson 
kinematics are reconstructed using the tight constrains of the decay 
$D^{*+} \rightarrow D^0\pi^+_{\rm slow} 
\rightarrow (K^-\pi^+)\pi^+_{\rm slow}$.
In order to make the comparison, the theory prediction for the 
charm production cross sections must be converted to those of  
charmed meson production.  This is done using a simple 
non-evolving Peterson {\em et al}.\ model \cite{pete} which depends on 
one parameter $\epsilon$ which is taken from $e^+e^-$ data.  The 
overall cross section normalization is set by the hadronization fraction 
$f(c \rightarrow D^*)$ again taken from $e^+e^-$ data.
The four-vector for the $D^*$ is constructed 
from that of the charmed quark by smearing the the charm three-vector 
in the lab frame.  The energy component is then fixed such that 
the four-vector has the physical $D^*$ mass, $2.01$ GeV.

Shown in Fig.~\ref{fig6} are $D^*$ meson cross sections measured 
by the H1 collaboration \cite{h1} differential in transverse 
momentum $p_{\perp}$, pseudo-rapidity $\eta$, reconstructed parton 
momentum fraction $x_g^{\rm obs}$, and momentum transfer $Q^2$ compared 
to the NLO calculation described in the Sec.~\ref{sec2} 
plus the Peterson {\it et al.} fragmentation model.  The shaded band 
corresponds to varying the charm quark mass of $1.5$ GeV by $\pm 0.2$ GeV.  
Overall the agreement is good, except for the pseudo-rapidity plot 
in which the theory under (over) estimates the data in the forward (backward)
region.

The ZEUS collaboration \cite{zeus} has also measured $D^*$ meson cross 
sections differential in momentum transfer $Q^2$, Bjorken $x$, hadronic 
energy $W$, transverse momentum $p_t$, pseudo-rapidity $\eta$, and $D^*$ 
momentum fraction $x(D^*)=2 | \vec{p}_{\gamma P{\rm cms}} | / W$ 
which are compared with theory in Fig.~\ref{fig7}.  
The boundaries of the bands correspond to 
varying the charm quark mass of $1.4$ GeV by $\pm 0.1$ GeV.
Again, the overall agreement is good, but the theory 
underestimates the data in the forward region and overestimates it 
in the backward region.  Additionally, 
the $D^*$ momentum fraction data, which is particularly sensitive to 
the charm hadronization process, is poorly described.  
Similar effects are seen in the $D^*$ photoproduction data 
at HERA.

Variations of the parton distribution set, 
renormalization/factorization scale, charm mass, or fragmentation parameter 
$\epsilon$ are unable to account for the differences between data and theory.
It also appears unlikely that an evolving fragmentation function would help; 
and the momentum transfers are large enough that any photon structure is  
surely negligible.

One explanation\cite{nor} proposed for the 
photoproduction data appears to work for the DIS data as well.  
Qualitatively, one is invited to think of a color 
string connecting the hadronizing charm quark and the proton remnant 
which pulls (drags) the charmed meson to the forward region.
This is made quantitative in the Lund String model modified for heavy 
flavor production \cite{lund}, as implemented in Pythia \cite{pythia}.
The shaded band in Fig.~\ref{fig7} shows what happens when the 
Peterson {\it et al.} model is replaced by an effective 
fragmentation model extracted from the Pythia based monte carlo 
RAPGAP\cite{rapgap}.  The agreement is much improved.

Another way to improve the agreement between data and theory 
is to simply raise the minimum $p_t$ of the events that are selected.  
Data from a slightly different decay chain, but higher 
minimum $p_t$ cut are shown as open triangles.  Here the Peterson and 
RAPGAP improved NLO predictions give essentially the 
same results, as expected.  However, a seemingly large fluctuation in the 
forward most data bin somewhat clouds this expectation.

A number of additional studies have been done.  For 
example, H1 \cite{h1}, using the above cross sections, has extracted a 
NLO gluon PDF which agrees well with their own gluon PDF, obtained 
indirectly through structure function scaling violations, and that of 
CTEQ4F3.  In a different approach, ZEUS \cite{zeus}, has extrapolated 
over the full phase space and extracted the structure function 
$F_2^{\rm charm}$.

In closing, the next-to-leading order calculations described herein have been 
very successful in describing charm production at HERA.  A variety of 
different observables have been studied, and the gluon parton 
distribution function 
and the charm contribution to proton structure function have been extracted.
The weakest stage of the calculation is, not surprisingly, 
modeling the hadronization 
of the produced charm to the observed charmed meson, especially at 
low transverse momentum.
One could take this as an opportunity to study hadronization 
in the presence of beam remnants.

\vspace*{4ex}
\noindent {\bf Acknowledgments.}  I thank the organizers 
for the invitation and Z. Sullivan for comments on the text.  
The work presented herein is the result of collaborations and 
discussions with J. Smith and E. Laenen, 
and was supported in part by the U.S.\ Department of Energy, High 
Energy Physics Division, Contract No.\ W-31-109-Eng-38.

%
\clearpage
\addcontentsline{toc}{section}{Index}
\flushbottom
\printindex

\end{document}